\documentstyle[12pt]{article}

\begin{document}

\newcommand{\dfrac}[2]{\displaystyle{\frac{#1}{#2}}}

{\it University of Shizuoka}

\hspace*{9.5cm} {\bf US-96-02}\\[-.3in]

\hspace*{9.5cm} {\bf AMU-96-01}\\[-.3in]

\hspace*{9.5cm} {\bf February 1996}\\[.3in]


\begin{center}

{\large\bf  A Democratic Seesaw Quark Mass Matrix }\\[.2in]

{\large\bf  Related to the Charged Lepton Masses }\footnote{
hep-ph/9602303}\\[.5in]

{\bf Yoshio Koide}\footnote{
E-mail: koide@u-shizuoka-ken.ac.jp} \\

Department of Physics, University of Shizuoka \\ 
395 Yada, Shizuoka 422, Japan \\[.1in]

and \\[.1in]

{\bf Hideo Fusaoka}\footnote{
E-mail: fusaoka@amugw.aichi-med-u.ac.jp} \\

Department of Physics, Aichi Medical University \\ 
Nagakute, Aichi 480-11, Japan \\[.5in]

{\large\bf Abstract}\\[.1in]

\end{center}

\begin{quotation}
We investigate a seesaw type mass matrix 
$M_f\simeq m_L M_F^{-1} m_R$  for quarks and leptons, $f$, 
under the assumptions that
the matrices $m_L$ and $m_R$ have common structures 
for the quarks and leptons, 
and that the matrix $M_F$ characterizing the heavy fermion sector
has the form [(unit matrix)+ (democratic-type matrix)].
We obtain well-satisfied relations for quark masses and mixings 
related to the charged lepton masses.
\end{quotation}

\newpage
\noindent
{\bf I. INTRODUCTION}

\vglue.05in

Why is the top quark mass $m_t$ so enhanced compared with the 
bottom quark mass $m_b$? Why is the $u$-quark mass $m_u$ of 
the order of the $d$-quark mass $m_d$\,?
In most models, in order to understand $m_t\gg m_b$, 
it is inevitable to bring in a parameter which takes hierarchically 
different values between up- and down-quark sectors.
However, from the point of view of the ``democracy of 
families", such a hierarchical difference seems to be unnatural.
What is of great interest to us is 
whether we can find a model in which $M_u$ and $M_d$ are almost 
symmetric in their matrix structures and in their 
parameter values.

Recently, by applying the  so-called ``seesaw" mechanism [1] 
to quark mass matrix  [2], the authors [3] have proposed a model
which provides explanations of both
$m_t \gg m_b$ and $m_u \sim m_d$, while keeping the model 
``almost" up-down symmetric. 
The essential idea is as follows: 
the mass matrices $M_f$ of quarks and leptons $f_i$ 
($i=1,2,3$: family index) are given by
$$
M_f \simeq -m_L M_F^{-1} m_R, \eqno(1.1)
$$
where $F_i$ denote heavy fermions $U_i$, $D_i$, 
$N_i$ and $E_i$, corresponding to $f_i = u_i, d_i, \nu_i$ 
and $e_i$, respectively. They have assumed that the mass 
matrix $m_L$ ($m_R$) between $f_L$ $(f_R)$ and $F_R$  $(F_L)$ 
is common to all $f = u, d, \nu, e$ 
(i.e., independently of up-/down- and quark-/lepton- sectors) 
and $m_R$ is proportional to $m_L$, i.e., $m_R=\kappa m_L$.
The variety of 
$M_f$  ($f = u, d, \nu, e)$ comes only from the variety 
of the heavy fermion matrix $M_F$  $(F = U, D, N, E)$. 
If we take a parametrization such that
the parameter value in the up-quark sector gives det$M_U\simeq 0$,
while, in down-quark sector, a value slightly deviated from that 
in $M_U$ does not yield det$M_D\simeq 0$ any longer, 
the model can provide $m_t \gg m_b$, keeping the model 
``almost" up-down symmetric, 
because of the factor $M_F^{-1}$ in the seesaw expression (1.1).
On the other hand, they have taken  $M_F=m_0 \lambda O_f$  as 
the form of the heavy fermion mass matrix $M_F$, where
$$
 O_f = \left(\begin{array}{ccc}
1 & 0 & 0 \\
0 & 1 & 0 \\
0 & 0 & 1 \end{array} \right) + b_f e^{i\beta_f} \left(
\begin{array}{ccc}
1 & 1 & 1 \\
1 & 1 & 1 \\
1 & 1 & 1 \end{array} \right) 
\equiv {\bf 1} + 3b_f e^{i\beta_f} X 
\ , \eqno(1.2)
$$
and $\lambda$ is an enhancement factor with $\lambda\gg \kappa \gg 1$.
Note that the inverse of the matrix $O_f$ is again given by 
the form [(unit matrix) + (democratic matrix)], i.e., 
$$
O_f^{-1} = {\bf 1} + 3a_f e^{i\alpha_f} X \ , \eqno(1.3) 
$$
with
$$
a_f e^{i \alpha_f} = -\dfrac{b_f e^{i\beta_f}}{1 + 3b_f e^{i\beta_f}} \ . 
\eqno(1.4)
$$
Thus, we can provide top-quark mass enhancement $m_t \gg m_b$ 
in the limit of $b_u e^{i \beta_u} \rightarrow -1/3$, 
because it leads to $|a_u| \rightarrow \infty$. On the 
other hand, since a democratic mass matrix [4] makes only one 
family heavy, we can keep $m_u \sim m_d$. 

They have taken 
$$
m_L = \dfrac{1}{\kappa} m_R = m_0 Z \equiv
 m_0 \left( \begin{array}{ccc}
z_1 & 0 & 0 \\
0 & z_2 & 0 \\
0 & 0 & z_3 \end{array} \right) \ , \eqno(1.5)
$$
where $z_i$ are normalized as $z_1^2+z_2^2+z_3^2=1$ and 
given by 
$$
\dfrac{z_1}{\sqrt{m_e}}=\dfrac{z_2}{\sqrt{m_\mu}}=
\dfrac{z_3}{\sqrt{m_\tau}}=\dfrac{1}{\sqrt{m_e+m_\mu+m_\tau}} \ , 
\eqno(1.6) 
$$
in order to give  the charged lepton mass matrix $M_e$ for
the case $b_e=0$,  i.e., $M_e = m_0 (\kappa/\lambda) Z^2$. 
They have obtained [3] reasonable quark mass ratios and 
Kobayashi-Maskawa (KM) [5] matrix parameters by taking 
$\kappa/\lambda=0.02$, 
$b_u=-1/3$, $\beta_u=0$, $b_d\simeq -1$ and $\beta\simeq -18^\circ$.

However, in their study [3], the KM matrix parameters have  
been evaluated only numerically, and have not
been expressed analytically in terms of charged lepton masses and 
mass matrix parameters $\kappa/\lambda$, $b_f$ and $\beta_f$.
Therefore, we cannot see  the dependencies of these parameters.
For example,  they predicted a value  $|V_{cb}| = 0.0598$, 
which is somewhat 
large compared with the recent experimental value [6] 
$|V_{cb}| = 0.041 \pm 0.003$.
However, we cannot see whether the discrepancy is a fatal defect 
in this model or not.
One of the purposes of the present paper is to express 
our predictions of $|V_{ij}|$ in terms of charged lepton mass ratios
and the mass matrix parameters $\kappa/\lambda$, $b_f$ and $\beta_f$.
We will obtain sum rules on 
quark mass ratios and $|V_{ij}|$, which are satisfied 
independently of those adjustable parameters.

Since they put stress on the ``economy of adjustable parameters" 
of the model, their predictions were done by adjusting 
only three parameters $\kappa/\lambda$, $b_d$ and $\beta_d$. 
As a result, some of the predictions were in poor agreement 
with experiment.
Another purpose of the present paper is to improve such 
disagreements by changing the model slightly.
We will be able to fit 
all the numerical predictions within experimental error.
Also, a possible shape of the unitary triangle 
$V_{ud} V_{ub}^{\ast} + V_{cd} V_{cb}^{\ast} 
+ V_{td} V_{tb}^{\ast} = 0$ 
in our model will be discussed. 

\vglue.2in
\noindent
{\bf II. A UNIVERSAL SEESAW MASS MATRIX WITH A DEMOCRATIC $M_F$}

\vglue.05in

In the present model, quarks and leptons $f_i$ belong to 
$f_L = (2,1)$ and 
$f_R = (1, 2)$ of SU(2)$_L \times $SU(2)$_R$ 
and heavy fermions $F_i$ are vector-like, i.e., 
$F_L = (1, 1)$ and $F_R = (1, 1)$. 
The SU(2)$_L$ and SU(2)$_R$ symmetries are broken by 
Higgs bosons $\phi_L=(\phi_L^+, \phi_L^0)$ and 
$\phi_R=(\phi_R^+, \phi_R^0)$, respectively.
We assume that these Higgs bosons 
couple to the fermions universally, but with the degree of freedom of 
their phases, as follows:
$$
H_{Yukawa}=\sum_{i=1}^3 (\overline{u} \ \overline{d})_{L i}
\left(y_{Li} \exp(i\delta^d_{Li})\right) \left(
\begin{array}{c}
\phi_L^+ \\
\phi_L^0 
\end{array}
\right) D_{Ri} $$
$$+  
\sum_{i=1}^3 (\overline{u} \ \overline{d})_{L i}
\left(y_{Li} \exp(i\delta^u_{Li})\right) \left(
\begin{array}{c}
\overline{\phi}_L^0 \\
-\phi_L^- 
\end{array}
\right) U_{Ri} 
$$
$$
+ h.c. +(L\leftrightarrow R) + [(u,d,U,D)\rightarrow (\nu,e,N,E)]
\ , \eqno(2.1)   
$$
where $y_{Li}$ and $y_{Ri}$ are real parameters, and they are 
universal for the quark and lepton sectors.
Therefore, the mass matrix  which is sandwiched by 
$(\overline{f}_L, \overline{F}_L)$  and $(f_R, F_R)^T$ is given by
a $6\times 6$ matrix  
$$
M = \left( \begin{array}{cc}
0 & m_L^f \\
m_R^{f\dagger} & M_F \end{array} \right) 
= m_0 \left( \begin{array}{cc}
0 & P_L^f Z \\
\kappa P_R^{f\dagger} Z & \lambda O_f \end{array} \right) \ , \eqno(2.2)
$$
where $m_L=y_{L_i}\langle \phi_L^0\rangle $, 
$P_L^f$ and $P_R^f$ are phase matrices given by 
$$
P_L^f = {\rm diag}(\exp(i\delta^f_{L1}), \exp(i\delta^f_{L2}), 
\exp(i\delta^f_{L3}))
\ , \eqno(2.3)
$$
and (2.3) with $(L\rightarrow R)$, and the matrices $Z$ and $O_f$ 
are defined by (1.5) and (1.2), respectively.
The previous model [3]  corresponds to the model (2.2) with 
$P_R^\dagger={\bf 1}$.

The KM matrix parameters are dependent only on
$$
P_{L}^{u\dagger} P_L^{d}\equiv P={\rm diag}(e^{i\delta_1}, 
e^{i\delta_2}, e^{i\delta_3}) \ .
\eqno(2.4)
$$
Of the three parameters $\delta_i$ ($i=1,2,3$), only two are observable.
Without loosing generality, we can put $\delta_1=0$. 
In the present model, the nine observable quantities 
(five  quark mass  ratios and four KM matrix parameters) are described by 
the seven parameters ($\kappa/\lambda$, 
$b_u$, $b_d$, $\beta_u$, $\beta_d$, $\delta_2$, $\delta_3$). 
Since we put the ansatz  ``maximal top-quark-mass enhancement" 
according to the Ref.~[3], we fix $b_u$ and $\beta_u$ at $b_u=-1/3$ and 
$\beta_u=0$.
However, we still possess five free parameters.
In order to economize on the number of the free parameters, we will give 
some speculation on these parameters in the final section.
On the other hand, since the phases $\delta_{L i}^e$ and 
$\delta_{R i}^e$ are not observable,
we can put $P_L^e=P_R^e={\bf 1}$.

\vglue.2in
\noindent
{\bf III. QUARK MASS RATIOS IN TERMS OF CHARGED LEPTON MASSES}

\vglue.05in

First, let us discuss the fermion mass spectra. Note that for the case 
$b_f = -1/3$ the seesaw expression (1.1) is not valid any longer
because of det$M_F=0$. 
In Fig.~1, we illustrate the numerical behavior of fermion masses 
$m_i^f$ versus the parameter $b_f$ which has been evaluated from the 
$6 \times 6 $ matrix (2.2)  without approximation.
As seen in Fig.~1, the third fermion is sharply 
enhanced at $b_f=-1/3$ for $\beta_f=0$. 
The calculation for the case $b_f\simeq -1/3$ must be done carefully.

For the case of $\lambda \gg \kappa \gg 1$, 
by expanding the eigenvalues $m_i^f$ ($i=1,2,3$) 
of the mass matrix (2.2) in $\kappa/\lambda$,  
we obtain the following expressions of $m_i^f$:
$$
\left(\dfrac{m_1^f}{m_0}\right)^2 = \dfrac{2\sigma^2}{\rho^2 f(b,\beta)}
\left( 1+\sqrt{1-\dfrac{4\sigma^2}{\rho^4} \dfrac{g(b,\beta)}{f^2(b,\beta)}}
\, \right)^{-1}
\left(\dfrac{\kappa}{\lambda}\right)^2 
+O\left(\dfrac{\kappa^4}{\lambda^4}\right) \ , \eqno(3.1)
$$
$$
\left(\dfrac{m_2^f}{m_0}\right)^2  = \dfrac{\rho^2 f(b,\beta)}{g(b,\beta)}
\left( 1+\sqrt{1-\dfrac{4\sigma^2}{\rho^4} \dfrac{g(b,\beta)}{f^2(b,\beta)}}\, 
\right)
\left( 1+\sqrt{1+4\rho^2 \dfrac{f(b,\beta)h(b,\beta)}{g^2(b,\beta)}}
\, \right)^{-1}
\left(\dfrac{\kappa}{\lambda}\right)^2 
$$
$$
+O\left(\dfrac{\kappa^4}{\lambda^4}\right) \ , \eqno(3.2)
$$
$$
\left(\dfrac{m_3^f}{m_0}\right)^2  = 3g(b,\beta)
\left[
\left(\dfrac{\kappa}{\lambda}\right)^2+6h(b,\beta)
\left( 1+\sqrt{1+4\rho^2 \dfrac{f(b,\beta)h(b,\beta)}{g^2(b,\beta)}}
\, \right)^{-1}
\right]^{-1} 
 \left(\dfrac{\kappa}{\lambda}\right)^2 
$$
$$
+O\left(\dfrac{\kappa^4}{\lambda^4}\right) \ , \eqno(3.3)
$$
where
$$
f(b,\beta)=(1+b)^2  - 2(1+2b) \dfrac{\sigma}{\rho^2} 
-4b\left(1-2\dfrac{\sigma}{\rho^2}\right)\sin^2\dfrac{\beta}{2}
\ , \eqno(3.4)
$$
$$
g(b,\beta)=(1+2b)^2  - 2(1+b)(1+3b)\rho -8b(1-2\rho)\sin^2\dfrac{\beta}{2}
\ , \eqno(3.5)
$$
$$
h(b,\beta)=(1+3b)^2  -12b\sin^2\dfrac{\beta}{2}
\ , \eqno(3.6)
$$
$$
\rho=z_1^2 z_2^2 + z_2^2 z_3^2 + z_3^2 z_1^2 \ , \eqno(3.7)
$$
$$
\sigma=z_1^2 z_2^2 z_3^2  \ , \eqno(3.8)
$$
and for simplicity we have denoted $b_f$ and $\beta_f$ as $b$ and 
$\beta$.
The explicit expressions of the up-quark masses at $b_u\simeq -1/3$ 
and the down-quark masses at $b_d\simeq -1$ are as follows:
$$
m_u \simeq \dfrac{3\sigma}{2\rho}\left( 
1+\dfrac{3\sigma}{4\rho^2} -\dfrac{3}{2}\varepsilon_u
\right) \dfrac{\kappa}{\lambda} m_0 
\simeq \dfrac{3m_e}{2m_\tau} \dfrac{\kappa}{\lambda} m_0 \ ,
\eqno(3.9)
$$ 
$$
m_c \simeq 2\rho \left[ 1-\dfrac{3\sigma}{4\rho^2}
-\dfrac{9}{2}\left(1-\dfrac{8}{3} \rho\right)\varepsilon_u  \right]
\dfrac{\kappa}{\lambda} m_0 \simeq 
2 \dfrac{m_\mu}{m_\tau} \dfrac{\kappa}{\lambda} m_0 \ , \eqno(3.10)
$$
$$
m_t \simeq \dfrac{1}{\sqrt{3}} \dfrac{1}{\sqrt{1 + 
27 \varepsilon_u^2 \lambda^2/\kappa^2}}
m_0 \simeq \dfrac{1}{\sqrt{3}}m_0 \ , \eqno(3.11)
$$
$$
m_d \simeq \dfrac{\sigma}{2|\sin(\beta_d/2)|\rho}
\left(1+\dfrac{1}{2}\varepsilon_d
\right) \dfrac{\kappa}{\lambda} m_0 \simeq 
\dfrac{1}{2|\sin(\beta_d/2)|}\dfrac{m_e}{m_\tau}\dfrac{\kappa}{\lambda} m_0 \ , 
\eqno(3.12)
$$
$$
m_s \simeq 2\left(1+\dfrac{3}{2}\varepsilon_d 
-2\sin^2\dfrac{\beta_d}{2}\right)
\left|\sin\dfrac{\beta_d}{2}\right|\, \rho \dfrac{\kappa}{\lambda} m_0 \simeq 
2\left|\sin\dfrac{\beta_d}{2}\right| \dfrac{m_\mu}{m_\tau}
\dfrac{\kappa}{\lambda} m_0 
\ , \eqno(3.13)
$$
$$
m_b \simeq \dfrac{1}{2}\left(1-\dfrac{1}{2}\varepsilon_d 
+ \dfrac{5}{2}\sin^2 \dfrac{\beta_d}{2}\right) \dfrac{\kappa}{\lambda} m_0 
\simeq \dfrac{1}{2} \dfrac{\kappa}{\lambda} m_0 \ , \eqno(3.14)
$$
where small parameters $\varepsilon_u$ and $\varepsilon_d$ 
are defined by
$$
\begin{array}{l}
b_u = -\dfrac{1}{3} + \varepsilon_u \ , \\
b_d = -1 + \varepsilon_d \ . 
\end{array} \eqno(3.15)
$$
Here, we have taken $\beta_u = 0$, because top-quark enhancement is 
caused only for the case of $\beta_u = 0$ (see Fig.~1). 
For down-quark masses, we have shown only the expressions for 
$b_d \simeq -1$ and $1 \gg \sin \beta_d \neq 0$, 
because from the numerical study in Ref.~[3], we have seen 
that the observed down-quark mass spectrum is in favor of 
$b_d \simeq -1$ and $\beta_d \simeq -20^{\circ}$. 

The expressions (3.9) -- (3.14) lead to the following sum rules 
which are almost independent of the parameters $\kappa/\lambda$, 
$\varepsilon_u$, $\varepsilon_d$ and $\beta_d$: 
$$
\dfrac{m_u}{m_c} \simeq \dfrac{3}{4} \dfrac{m_e}{m_\mu} \ , 
\eqno(3.16)
$$
$$
\dfrac{m_c}{m_b} \simeq 4 \dfrac{m_\mu}{m_\tau}\ , 
\eqno(3.17)
$$
$$
\dfrac{m_d m_s}{m_b^2} \simeq 4 \dfrac{m_e m_\mu}{m_\tau^2} \ , 
\eqno(3.18)
$$
$$
\dfrac{m_u}{m_d} \simeq 3 \dfrac{m_s}{m_c} \simeq \dfrac{3}{4} 
\dfrac{m_d}{m_b} \dfrac{m_\tau}{m_\mu} \simeq 
3\left|\sin\dfrac{ \beta_d}{2}\right| \ . 
\eqno(3.19)
$$
The expression (3.16) has been obtained by the model 
$M_u \propto Z O^{-1}_f Z$ in Ref.[7]. 

In the limit of unbroken SU(2)$_L \times $SU(2)$_R$, i.e., 
$m_L=m_R=0$, heavy fermion masses $m_{F'_i}$ are given by 
$$
\begin{array}{ll}
m_{F'_1}=m_{F'_2}= \lambda m_0\ , \\[.2in]
m_{F'_3}=\sqrt{1+6 b_f \cos\beta_f +9b_f^2} \lambda m_0 \ ,
\end{array} \eqno(3.20)
$$ 
where $F'_i$ are mass-eigenstates for the mass matrix $M_F = m_0 \lambda O_f$ 
given by (1.2). As seen from (3.20), the minimum condition of the sum of 
the up-heavy-quark masses leads to $\beta_u = 0$ and $b_u = -1/3$. 
Therefore, the ansatz ``maximal top-quark-mass enhancement" can 
be replaced by another expression that the parameters $(b_u, \beta_u)$ 
are fixed such that the sum of the up-heavy-quark masses becomes a minimum. 

For the case of $Z\neq 0$, the heavy fermion masses are given by 
$$
m_4^e \simeq m_5^e \simeq m_6^e \simeq \lambda m_0 \ , \eqno(3.21) 
$$
$$
m_4^u \simeq \dfrac{1}{\sqrt{3}} \kappa m_0, \ \ \ 
m_5^u \simeq m_6^u \simeq \lambda m_0 \ , 
\eqno(3.22)
$$
$$
m_4^d \simeq m_5^d \simeq \lambda m_0, \ \ \ 
m_6^d \simeq 2 \sqrt{1 + 3 \sin^2(\beta_d/2)} \lambda m_0 \ , \eqno(3.23)
$$
where the numbering of $m_i^f$ has been defined as 
$m_4^f \leq m_5^f \leq m_6^f$ in the mass eigenstates 
$F'_i$ $(i = 1, 2, 3)$. 
Note that only the fourth up-quark 
$u_4$ ($\equiv U'_3$) is remarkably light compared with 
other heavy fermions.  
The enhancement of the top-quark $u_3$ ($\equiv t$) is caused 
at the cost of the lightening of $U'_3$. Since the 
mass ratio $m_4^u/m_3^u$ is given by 
$$
m_4^u/m_t \simeq \kappa \eqno(3.24)
$$
and $\kappa$ is of the order of $m(W_R)/m(W_L)$, we can expect 
the observation of the fourth up-quark $u_4$ at an energy 
scale at which we can observe the right-handed weak bosons 
$W_R$.


\vglue.2in
\noindent
{\bf IV. KM MATRIX PARAMETERS IN TERMS OF CHARGED LEPTON MASSES}

\vglue.05in

We diagonalize the $6\times 6$ mass matrix (2.2) by the following 
two steps. 
As the first step, we transform the mass matrix $M$ (2.2) into 
$$
M'=\left(
\begin{array}{cc}
M'_{11} & 0 \\
0 & M'_{22} 
\end{array}\right) 
\equiv \left(
\begin{array}{cc}
M_{f} & 0 \\
0 & M'_{F} 
\end{array}\right) 
\ . \eqno(4.1)
$$
At the second step, 
the $3\times 3$ matrix $M_f\equiv M'_{11}$ with $P_L^f=P_R^f={\bf 1}$,
i.e., $\widetilde{M}_f\equiv P_L^{f\dagger} M_f P_R^f$, is diagonalized 
by two unitary matrices $U_L^f$ and $U_R^f$ as
$$
 U_L^f \widetilde{M}_f U_R^{f\dagger} = D_f \ , \eqno(4.2)
$$
where $D_f={\rm diag}(m_1^f, m_2^f, m_3^f)$.
The KM matrix $V$ is given by 
$$
V=U_L^u P U_L^{d\dagger} \ , \eqno(4.3)
$$
where the phase matrix $P$ is defined by (2.4).

For the up-quark sector, 
we put an ansatz ``maximal top quark mass enhancement",
i.e., we assume that $b_u=-1/3$ and $\beta_u=0$. 
Then, the unitary matrix $U_L^u$ is given by 

\renewcommand{\arraystretch}{2}
$$
U_L^u \simeq \left( 
\begin{array}{ccc} 
\displaystyle 
1 & 
-\displaystyle \dfrac{1}{2} \dfrac{z_1}{z_2} 
& \displaystyle -\dfrac{1}{2} \dfrac{z_1}{z_3} \\
\displaystyle \dfrac{1}{2} \dfrac{z_1}{z_2} & 1 & 
\displaystyle - \dfrac{z_2}{z_3} \\
\displaystyle \dfrac{z_1}{z_3} 
& \displaystyle \dfrac{z_2}{z_3} & 1
\end{array} \right) \ , \eqno(4.4)
$$
For the down-quark sector, we use the following approximate expression 
for $b_d=-1$, 
$$
U_L^d \simeq
\left( 
\begin{array}{ccc}
1 & \displaystyle - \dfrac{z_1}{z_2}
    \dfrac{ie^{i\beta_d/2}}{2\sin(\beta_d/2)}
  & \displaystyle -\dfrac{z_1}{z_3} 
  \dfrac{ie^{i\beta_d/2}}{2\sin(\beta_d/2)}\\
\displaystyle - \dfrac{z_1}{z_2}\dfrac{ie^{-i\beta_d/2}}{2\sin(\beta_d/2)} 
 & 1 
 & \displaystyle  \dfrac{z_2}{z_3} \dfrac{2-e^{i\beta_d}}{5-4\cos\beta_d} \\ 
\displaystyle -\dfrac{z_1}{z_3}\dfrac{2-e^{-i\beta_d}}{5-4\cos\beta_d} 
 & \displaystyle -\dfrac{z_2}{z_3}\dfrac{2-e^{-i\beta_d}}{5-4\cos\beta_d} 
 & 1
\end{array} \right) \ . \eqno(4.5)
$$
\renewcommand{\arraystretch}{1}
Here, the expression (4.5) is valid only for a sizable value of $\beta_d$,
i.e., for $(z_1/z_3)^2 < \beta_d^2 \ll 1$. 

Without losing  generality, we can take 
$$
P={\rm diag}(1,e^{i\delta_2}, e^{i\delta_3}) \ , \eqno(4.6)
$$
so that we obtain the following KM matrix elements:
$$
V_{12}\simeq \dfrac{z_1}{2z_2}\left( \dfrac{i}{\sin\dfrac{\beta_d}{2}}
e^{i\beta_d/2}
-e^{i\delta_2}\right) \ , \eqno(4.7)
$$
$$
V_{23}\simeq -\dfrac{z_2}{z_3}\left( \dfrac{ 2-e^{i\beta_d}}{5-4\cos\beta_d}
 e^{i\delta_2}+e^{i\delta_3}\right) \ ,\eqno(4.8)
$$
$$
V_{13}\simeq -\dfrac{z_1}{2z_3} \left[ \dfrac{2-e^{i\beta_d}}{5-4\cos\beta_d}
\left( 2-e^{i\delta_2}\right) +e^{i\delta_3} \right] \ , \eqno(4.9)
$$
$$
V_{31}\simeq \dfrac{z_1}{z_3}\left[ 1
+\dfrac{i e^{-i \beta_d /2}} {2\sin\dfrac{\beta_d}{2}}
\left( e^{i\delta_2} + e^{i\delta_3}\right) \right] \ . \eqno(4.10)
$$

Eq.~(4.7) leads to 
$$
|V_{us}|\simeq \dfrac{z_1}{2z_2}\dfrac{1}{|\sin\dfrac{\beta_d}{2}|}
\sqrt{1+2\sin\dfrac{\beta_d}{2}\sin\left(\dfrac{\beta_d}{2}-\delta_2\right)
+\sin^2\dfrac{\beta_d}{2} } \ . \eqno(4.11)
$$
If we assume  $|\delta_2|\ll |\beta_d|\ll 1$, we obtain the well-known 
formula [8]
$$
|V_{us}|\simeq \sqrt{{m_d}/{m_s}} \ , \eqno(4.12)
$$
from (3.12) and (3.13).

Since we have already known that $\sin^2 (\beta_d/2)\simeq 0.025$ 
from the observed value of
$m_s/m_c$ and $z_2/z_3\simeq 0.24$ 
from the observed value of $m_\mu/m_\tau$, 
we must take $\delta_3-\delta_2\simeq \pi$ in order to understand 
the observed value $|V_{cb}|\simeq 0.041$ [6].
When we put $\delta=\delta_3-\delta_2 -\pi$ ($|\delta|\ll 1$), 
we obtain
$$
|V_{cb}|\simeq 2\dfrac{z_2}{z_3}
\dfrac{\left| \sin\dfrac{\beta_d}{2}+\sin\dfrac{\delta}{2}\right|}{
\sqrt{5-4\cos\beta_d} } \simeq 
 \dfrac{m_s}{\sqrt{m_c m_b}}
\dfrac{\left| 1+{\sin\dfrac{\delta}{2}}/{\sin\dfrac{\beta_d}{2}}\right|}{
\sqrt{1+8\sin^2\dfrac{\beta_d}{2}} } \ , \eqno(4.13)
$$
where we have used (3.10), (3.13) and (3.14).

Similarly, we obtain
$$
|V_{ub}|\simeq \dfrac{z_1}{z_3}
\dfrac{\left| \sin\dfrac{\beta_d}{2}+\sin\dfrac{\delta}{2}
+2\sin\dfrac{\delta_2}{2}\right|}{\sqrt{5-4\cos\beta_d}}
 \ , \eqno(4.14)
$$
so that 
$$
\left| \dfrac{V_{ub}}{V_{cb}}\right|\simeq 
\dfrac{z_1}{2z_2}\left| 1+
\dfrac{2\sin\dfrac{\delta_2}{2}}{\sin\dfrac{\beta_d}{2}
+\sin\dfrac{\delta}{2}} \right|
\simeq \sqrt{\dfrac{m_u}{2m_c}}\left| 1+
\dfrac{2\sin\dfrac{\delta_2}{2}}{\sin\dfrac{\beta_d}{2}
+\sin\dfrac{\delta}{2}} \right|
 \ , \eqno(4.15)
$$
or
$$
\left| \dfrac{V_{ub}}{V_{cb}}\right|\simeq 
|V_{us}|\,\left|\sin\dfrac{\beta_d}{2}\right| \left( 1+
\dfrac{2\sin\dfrac{\delta_2}{2}}{\sin\dfrac{\beta_d}{2}
+\sin\dfrac{\delta}{2}} \right)
 \ . \eqno(4.16)
$$
For $|V_{td}|$, we obtain
$$
|V_{td}| \simeq \dfrac{z_1}{z_3}\left| 1 +
\dfrac{\sin\dfrac{\delta}{2}}
{\sin\dfrac{\beta_d}{2}}\right| \ , \eqno(4.17)
$$
or
$$
\left| \dfrac{V_{td}}{V_{cb}}\right|\simeq |V_{us}|
\sqrt{
\dfrac{1+8\sin^2 \dfrac{\beta_d}{2} }{1+\sin^2\dfrac{\beta_d}{2}
+2\sin\dfrac{\beta_d}{2}\sin\left(\dfrac{\beta_d}{2}-\delta_2\right)} 
}
\ . \eqno(4.18)
$$

The rephasing invariant $J$ [9] is expressed in terms of $|V_{ij}|$ 
as follows [10]:
$$
J^2 = |V_{us}|^2 |V_{cb}|^2 |V_{ub}|^2 \left(
1+|V_{us}|^2-|V_{cb}|^2-\omega\right)
$$
$$
-\dfrac{1}{4}\left[ |V_{us}|^2 |V_{cb}|^2-\left(|V_{us}|^2+|V_{cb}|^2
\right) |V_{ub}|^2 +\left( 1-|V_{ub}|^2\right) \omega \right]^2 \ , 
\eqno(4.19)
$$
where
$$
\omega =|V_{cd}|^2-|V_{us}|^2=|V_{ts}|^2-|V_{cb}|^2
=|V_{ub}|^2-|V_{td}|^2
\ . \eqno(4.20)
$$
By using (4.18), i.e., $|V_{td}|^2\simeq |V_{us}|^2 |V_{cb}|^2$, and 
the observed relation $|V_{us}|^2\gg |V_{cb}|^2 \gg |V_{ub}|^2$, 
we obtain
$$
|J|\simeq \sqrt{
1-\dfrac{1}{4} \dfrac{|V_{ub} /V_{cb}|^2}{|V_{us}|^2 }}
|V_{us}|\, |V_{cb}|\, |V_{ub}| \ .
\eqno(4.21)
$$

\vglue.2in
\noindent
{\bf V. NUMERICAL RESULTS OF THE KM MATRIX PARAMETERS}

\vglue.05in

In the previous section, we have obtained  approximate expressions 
for the KM matrix elements $|V_{ij}|$.
The results for $|V_{us}|$, $|V_{cb}|$ and $|V_{td}|$, i.e., (4.11), 
(4.13) and (4.17) are in excellent agreement with the results from 
numerical evaluation of the $6\times 6$ mass matrix (2.2). 
However, for the matrix element $|V_{ub}|$, the numerical value 
of (4.14) is somewhat in disagreement with that 
which is directly evaluated from the diagonalization of the 
$6\times 6$ mass matrix (2.2).
This means that the approximate expressions (4.4) and (4.5) are not 
sufficient to evaluate a small mixing element such as $|V_{ub}|$.
However, the expression (4.14) is still useful for describing 
the dominant behavior of $|V_{ub}|$.

In order to complement the study of the previous section, 
in the present section, we shall present a numerical study of 
the $6\times 6$ mass matrix (1.4).
The results for the mass eigenstates given in Sec.~III are valid with 
good accuracy, so we confine our numerical study to that of the 
KM matrix parameters (for the numerical study of the mass eigenvalues, 
see Ref.~[3]). 

As the numerical inputs, according to Ref.~[3], 
we use $\kappa/\lambda=0.02$, $b_u=-1/3$, $\beta_u=0$, $b_d=-1$ and 
$\beta_d=-18^\circ$, which are required for a reasonable 
fit with the observed quark masses.
Our interest is in the behavior of $|V_{ij}|$ versus the phase 
parameters $\delta_2$ and $\delta_3$ defined by (2.4),
because in the previous study [3], the degree of freedom of the phases 
$(\delta_2, \delta_3)$ was not taken into consideration.
In Fig.~2, we illustrate the allowed regions of $(\delta_2, \delta_3)$ 
which give the observed values of $|V_{us}|$, $|V_{cb}|$ and 
$|V_{ub}|$ [6]:
$$
\begin{array}{l}
|V_{us}|= 0.2205 \pm 0.0018 \ , \\ 
|V_{cb}|=  0.041\pm 0.003 \ , \\
|V_{ub}/V_{cb}|= 0.08\pm 0.02 \ .
\end{array} \eqno(5.1)
$$
We have two allowed regions of $(\delta_2,\delta_3)$: 
we obtain the predictions 
$$
|V_{us}|= 0.2195 \ , \ \ \ |V_{cb}|= 0.0388  \ , 
\ \ \ |V_{ub}|=0.0028 \ ,
$$
$$
|V_{ub}/V_{cb}|= 0.072  \ , \ \ \ |V_{td}|=0.0105 \ , 
\ \ \ J= -1.7\times 10^{-5} \   ,
\eqno(5.2)
$$
for $(\delta_2,\delta_3)=(0^\circ , 186^\circ )$ and
$$
|V_{us}|= 0.2211 \ , \ \ \ |V_{cb}|= 0.0411 \ , \ \ \ |V_{ub}|=0.0027 \ ,
$$
$$
|V_{ub}/V_{cb}|= 0.065 \ , \ \ \ |V_{td}|=0.0092  \ , 
\ \ \ J= -2.3\times 10^{-5}  \   ,
\eqno(5.3)
$$
for $(\delta_2,\delta_3)=(4^\circ , 208^\circ )$. In the latter case 
($\delta_2 = 4^\circ$), we have taken such a value of $\delta_3$ at which 
the Wolfenstein parameter [11] $\rho$ [which is defined by $V_{ub} \equiv 
|V_{us}||V_{cb}|(\rho - i \eta)$] takes $\partial \rho/\partial\delta_3 = 0$ 
by way of trial.

In Fig.~3, we show the possible unitary-triangle shape of the present 
model on the $(\rho , \eta)$ plane. 
The vertex $(\rho, \eta)$ moves on the circle which is denoted by 
the solid line in Fig.~3 according as the parameter $\delta_3$ varies from 
$0^\circ$ to $360^\circ$.
For reference, we have shown the constraints [12] from the observed values 
$|V_{ub}/V_{cb}|$, $\Delta m_{B_d}$ and $\varepsilon_K$.
Both triangles which correspond to the cases 
$(\delta_2, \delta_3) = (0^\circ, 186^\circ)$ and 
$(4^\circ, 208^\circ)$ satisfy these constraints safely.

\vglue.2in
\noindent
{\bf VI. DISCUSSIONS}

\vglue.05in

In conclusion, we have obtained relations among quark mass ratios and 
KM matrix parameters  
on the basis of the democratic seesaw mass matrix (2.2). 
The sum rules given in III and IV are well satisfied  by the 
observed values.

In the present model, there are seven parameters.
Two of the seven, $(b_u, \beta_u)$, have been fixed as 
$(b_u, \beta_u)=(-1/3,0)$ by putting the ansatz ``maximal 
top-quark-mass enhancement" (or ``minimal up-heavy-quark masses").
The values $(b_d, \beta_d)$ have been fixed as 
$(b_d, \beta_d)\simeq (-1,-18^\circ)$ from the phenomenological study [3].
Why does the parameter $b_f$ take $b_e=0$, $b_u=-1/3$ and $b_d=-1$? 

If we consider an SU(3)-family symmetry, the parameter $b_f e^{i\beta_f}$ 
gives a measure of its symmetry breaking.
The symmetry is exactly unbroken for the charged heavy leptons $E_i$, 
i.e., $m_{E_1}=m_{E_2}=m_{E_3}= \lambda m_0$, in the limit of $m_L=m_R=0$.
For up-heavy-quarks, the symmetry is badly broken, i.e., 
$m_{U_1}=m_{U_2}= \lambda m_0$ and $m_{U_3}=0$.
If the values $(b_f, \beta_f)$ are governed by a rule, the rule should be 
independent of low energy phenomena, i.e., 
the values $(b_f, \beta_f)$ should be determined only by the dynamics 
of heavy fermions $F_i$, 
independently of that of quarks and leptons $f_i$.
For example, let us direct our attention to the deviation 
$$
\Delta m_F=m_{F_3}-m_{F_2}=m_{F_3}-m_{F_1}
=\left( \sqrt{1+6 b_f \cos\beta_f + 9 b_f^2} - 1 \right) \lambda m_0 \ , 
\eqno(6.1)
$$
which is derived from (3.20). 
If we assume that $\Delta m_U + \Delta m_D =0$, 
then we obtain
$$
b_d=-\dfrac{1}{3}\left( 1-2\sin^2\dfrac{\beta_d}{2} + 
2 \sqrt{1-\dfrac{1}{2}\sin^2\dfrac{\beta_d}{2} }\right) \ , \eqno(6.2)
$$
which gives $b_d\simeq -1$ for $|\beta_d|\ll 1$.
The similar ansatz $\Delta m_E + \Delta m_N =0$, i.e., $\Delta m_N=0$, 
predicts $b_\nu=0$ or 
$$
b_\nu = - \dfrac{2}{3}\cos\beta_\nu \ . \eqno(6.3)
$$
The latter solution  predicts $b_\nu \simeq -2/3$ for $|\beta_\nu|\ll 1$.

Another interesting speculation on $b_f$ is as follows:
If we plot the values of $b_f$ and the electric charges $Q_f$ on the 
$(b_f, Q_f)$ plane, the points $(b_e, Q_e)=(0,-1)$,  $(b_d, Q_d)=(-1,-1/3)$, 
$(b_u, Q_u)=(-1/3,+2/3)$ take three corners of a square on $(b_f, Q_f)$.
The remaining corner is assigned to $(b_\nu, Q_\nu)=(+2/3, 0)$.
In other words, the parameter $b_f$ is given by an empirical relation
$$
\dfrac{3}{2} b_f = Q_f - \dfrac{1}{2} N_B + (N_L - 3 N_B) \ , \eqno(6.4)
$$
where $N_L$ and $N_B$ are lepton- and baryon-numbers, respectively.

Whether these speculations are justified or not will be checked by seeing 
whether neutrino masses and mixings can be described by a similar model 
with $b_\nu \simeq \pm 2/3$. 
A study of neutrino mixings based on the democratic seesaw mass matrix model 
will be given elsewhere [13].

\vglue.2in

\centerline{\bf ACKNOWLEDGMENTS}

The authors would like to express their sincere thanks to Professors 
R.~Mohapatra, K.~Hagiwara and M.~Tanimoto for  their valuable comments 
on an earlier version of the present work.
This work was supported by the Grant-in-Aid for Scientific Research, the 
Ministry of Education, Science and Culture, Japan (No.06640407). 



\newpage

\vglue.3in
\newcounter{0000}
\centerline{\bf REFERENCES AND FOOTNOTES}
\begin{list}
{[~\arabic{0000}~]}{\usecounter{0000}
\labelwidth=0.8cm\labelsep=.1cm\setlength{\leftmargin=0.7cm}
{\rightmargin=.2cm}}

\item M.~Gell-Mann, P.~Rammond and R.~Slansky, in {\it Supergravity}, 
edited by P.~van Nieuwenhuizen and D.~Z.~Freedman (North-Holland, 
1979); 
T.~Yanagida, in {\it Proc. Workshop of the Unified Theory and 
Baryon Number in the Universe}, edited by A.~Sawada and A.~Sugamoto 
(KEK, 1979); 
R.~Mohapatra and G.~Senjanovic, Phys.~Rev.~Lett.~{\bf 44}, 912 (1980).

\item Z.~G.~Berezhiani, Phys.~Lett.~{\bf 129B}, 99 (1983);
Phys.~Lett.~{\bf 150B}, 177 (1985);
D.~Chang and R.~N.~Mohapatra, Phys.~Rev.~Lett.~{\bf 58}, 1600 (1987); 
A.~Davidson and K.~C.~Wali, Phys.~Rev.~Lett.~{\bf 59}, 393 (1987);
S.~Rajpoot, Mod.~Phys.~Lett. {\bf A2}, 307 (1987); 
Phys.~Lett.~{\bf 191B}, 122 (1987); Phys.~Rev.~{\bf D36}, 1479 (1987);
K.~B.~Babu and R.~N.~Mohapatra, Phys.~Rev.~Lett.~{\bf 62}, 1079  (1989); 
Phys.~Rev. {\bf D41}, 1286 (1990);  
S.~Ranfone, Phys.~Rev.~{\bf D42}, 3819 (1990); 
A.~Davidson, S.~Ranfone and K.~C.~Wali, Phys.~Rev.~{\bf D41}, 208 (1990); 
I.~Sogami and T.~Shinohara, Prog.~Theor.~Phys.~{\bf 66}, 1031 (1991);
Phys.~Rev.~{\bf D47}, 2905 (1993); 
Z.~G.~Berezhiani and R.~Rattazzi, Phys.~Lett.~{\bf B279}, 124 (1992);
P.~Cho, Phys.~Rev.~{\bf D48}, 5331 (1994); 
A.~Davidson, L.~Michel, M.~L,~Sage and  K.~C.~Wali, Phys.~Rev.~{\bf D49}, 
1378 (1994); 
W.~A.~Ponce, A.~Zepeda and R.~G.~Lozano, Phys.~Rev.~{\bf D49}, 4954 (1994).
\item Y.~Koide and H.~Fusaoka, US-95-03 \& AMU-95-04 (1995) 
(hep-ph/9505201), to be published in Z.~Phys.~C.
\item H.~Harari, H.~Haut and J.~Weyers, Phys.~Lett.~{\bf 78B}, 459 (1978);
T.~Goldman, in {\it Gauge Theories, Massive Neutrinos and 
Proton Decays}, edited by A.~Perlumutter (Plenum Press, New York, 
1981), p.111;
T.~Goldman and G.~J.~Stephenson,~Jr., Phys.~Rev.~{\bf D24}, 236 (1981); 
Y.~Koide, Phys.~Rev.~Lett. {\bf 47}, 1241 (1981); 
Phys.~Rev.~{\bf D28}, 252 (1983); {\bf 39}, 1391 (1989);
C.~Jarlskog, in {\it Proceedings of the International Symposium on 
Production and Decays of Heavy Hadrons}, Heidelberg, Germany, 1986
edited by K.~R.~Schubert and R. Waldi (DESY, Hamburg), 1986, p.331;
P.~Kaus, S.~Meshkov, Mod.~Phys.~Lett.~{\bf A3}, 1251 (1988); 
Phys.~Rev.~{\bf D42}, 1863 (1990);
L.~Lavoura, Phys.~Lett.~{\bf B228}, 245 (1989); 
M.~Tanimoto, Phys.~Rev.~{\bf D41}, 1586 (1990);
H.~Fritzsch and J.~Plankl, Phys.~Lett.~{\bf B237}, 451 (1990); 
Y.~Nambu, in {\it Proceedings of the International Workshop on 
Electroweak Symmetry Breaking}, Hiroshima, Japan, (World 
Scientific, Singapore, 1992), p.1.
\item M.~Kobayashi and T.~Maskawa, Prog.~Theor.~Phys.~{\bf 49}, 652 (1973).
\item Particle data group, Phys.~Rev.~{\bf D50}, 1173 (1994) and 
1995 off-year partial update for the 1996 edition available on the PDG 
WWW pages (URL: http://pdg.lbl.gov/).
\item Y.~Koide, Mod.~Phys.~Lett.~{\bf A8}, 2071 (1993).
\item S.~Weinberg, Ann.~N.Y.~Acad.~Sci.~{\bf 38}, 1945 (1977); 
H.~Fritszch, Phys.~Lett. {\bf 73B}, 317 (1978); 
Nucl.~Phys.~{\bf B155}, 189 (1979) ;
H.~Georgi and D.~V.~Nanopoulos, {\it ibid}.~{\bf B155}, 52 (1979).
\item C.~Jarlskog, Phys.~Rev.~Lett.~{\bf 55} (1985)  1839; 
O.~W.~Greenberg, Phys.~Rev. {\bf D32}, 1841 (1985); 
I.~Dunietz, O.~W.~Greenberg, and D.-d.~Wu, Phys.~Rev.~Lett. {\bf 55}, 
2935 (1985); 
C.~Hamzaoui and A.~Barroso, Phys.~Lett.~{\bf 154B}, 202 (1985); 
D.-d.~Wu, Phys.~Rev.~{\bf D33}, 860 (1986).
\item C.~Hamzaoui, Phys.~Rev.~Lett.~{\bf 61}, 35 (1988); 
G.~C.~Branco and L.~Lavoura, Phys.~Lett.~{\bf B208}, 123 (1988).
\item L.~Wolfenstein, Phys.~Rev.~{\bf D45}, 4186 (1992).
\item For instance, see A.~Pich and J.~Prades, Phys.~Lett.~{\bf B346}, 342 
(1995).
\item The work has, in part, been reported in Y.~Koide, 
the IV International Symposium on {\it Weak and Electromagnetic 
Interactions in Nuclei}, June 12 - 16, 1995, Osaka, Japan, and 
Y.~Koide, 
Report No. US-95-07 (1995) (hep-ph/9508369).  
\end{list}

\newpage

\begin{center}
{\bf Figure Captions}
\end{center}

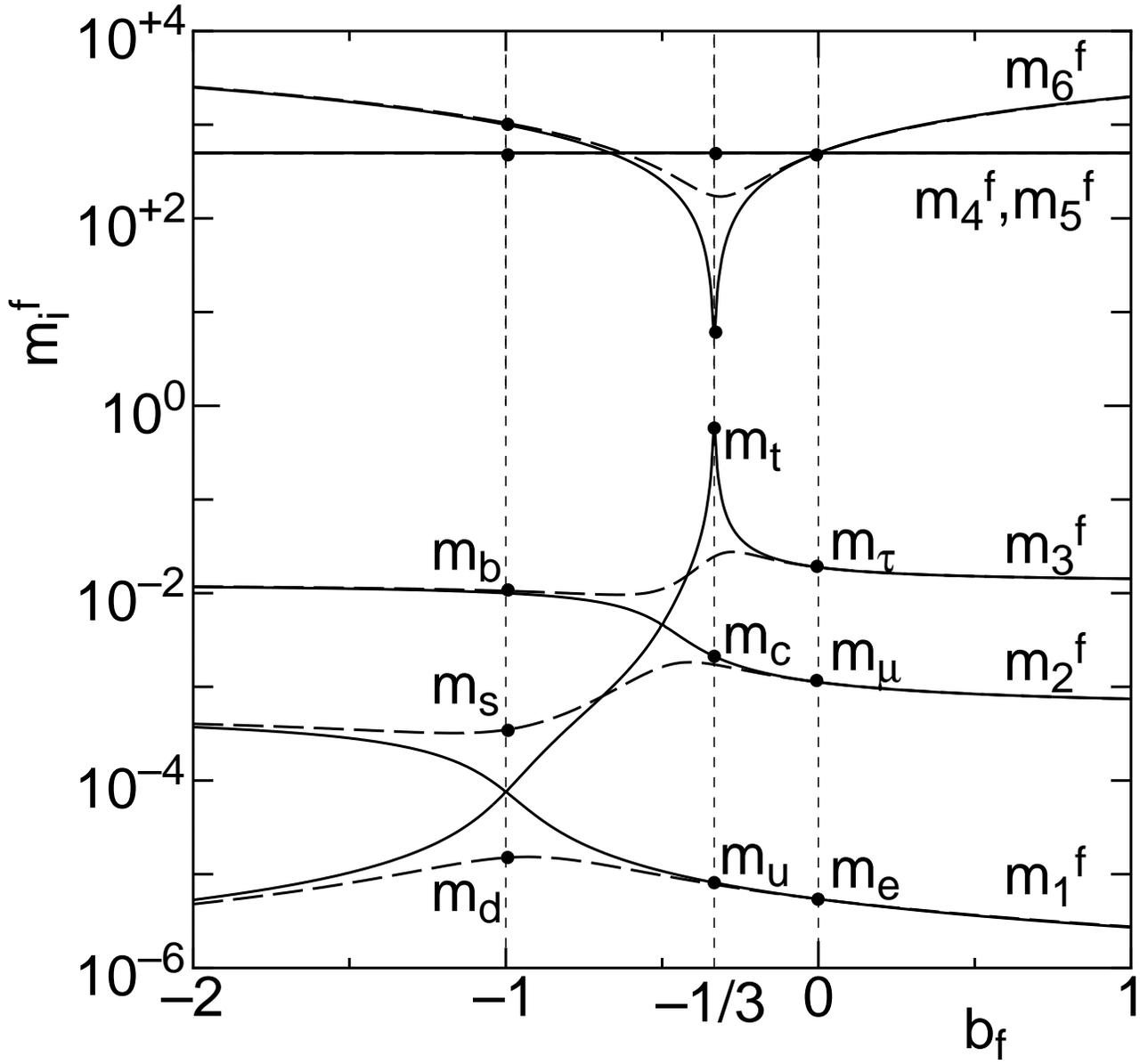
Fig.~1. Masses  $m_i^f$ ($i=1,\cdots,6$) versus $b_f$ for the case of 
$\kappa=10$ and  $\kappa/\lambda=0.02$.
The solid and broken lines denote for the cases of $\beta_f=0$ 
and $\beta_f=-20^\circ$, respectively. 
The parameters $\kappa$ and $\lambda$ are defined by (2.2). 
At $b_f=0$, the charged lepton masses $m_e$, $m_\mu$ and $m_\tau$ 
have been used as input values for the parameters $z_i$.
For up- and down-quark sectors, the values $b_u=-1/3$ and $b_d=-1$ 
are chosen from the phenomenological study [3] of the observed 
quark masses.

\vglue.1in

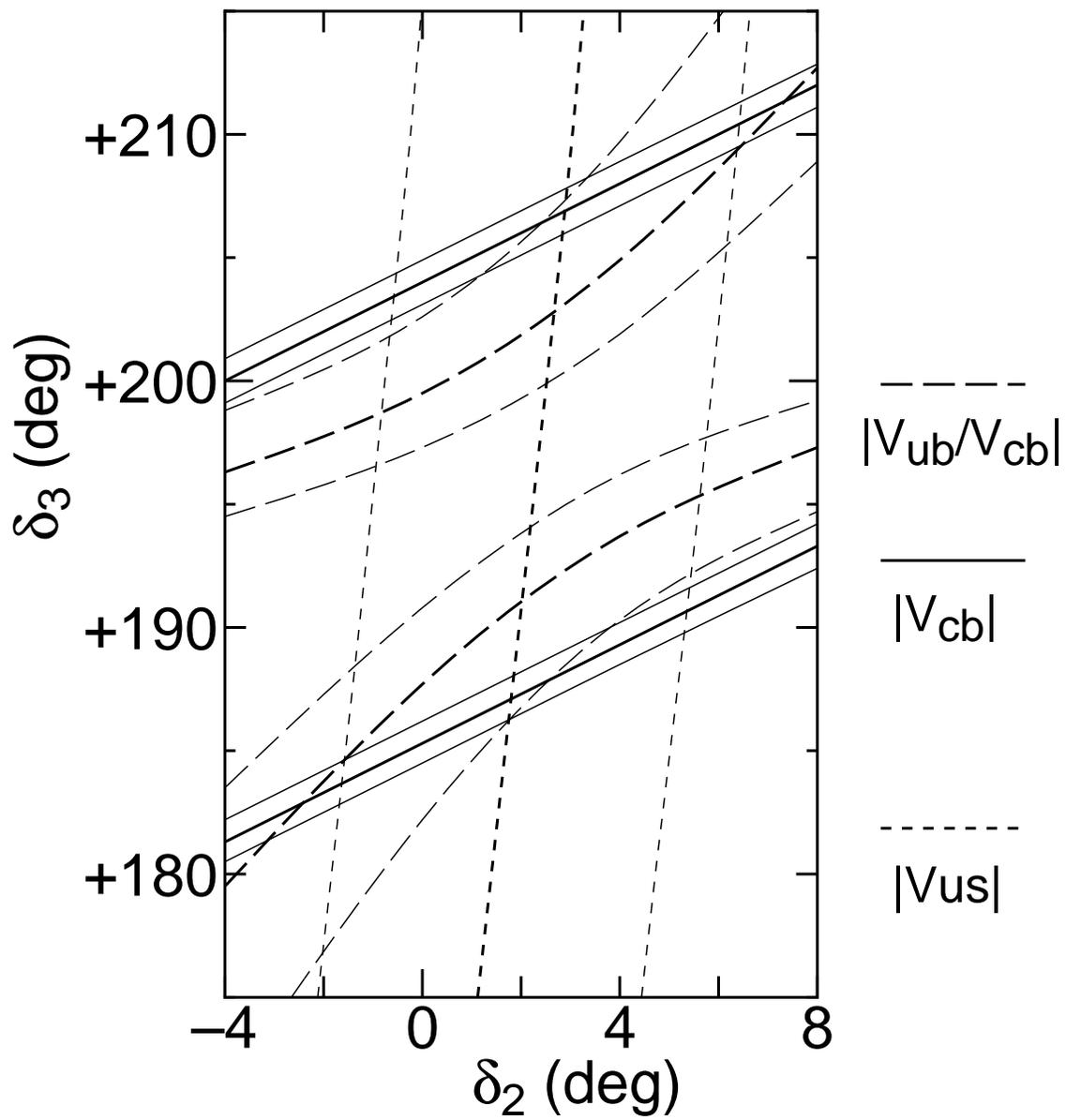
Fig.~2.  
Constraints on the phase parameters $(\delta_2, \delta_3)$ from 
the experimental values $|V_{us}|=0.2205\pm 0.0018$ (dotted lines), 
$|V_{cb}|=0.041\pm 0.003$ (solid lines) and 
$|V_{ub}/V_{cb}|=0.08\pm 0.02$ (dashed lines).
The hatched areas denote the allowed regions.

\vglue.1in

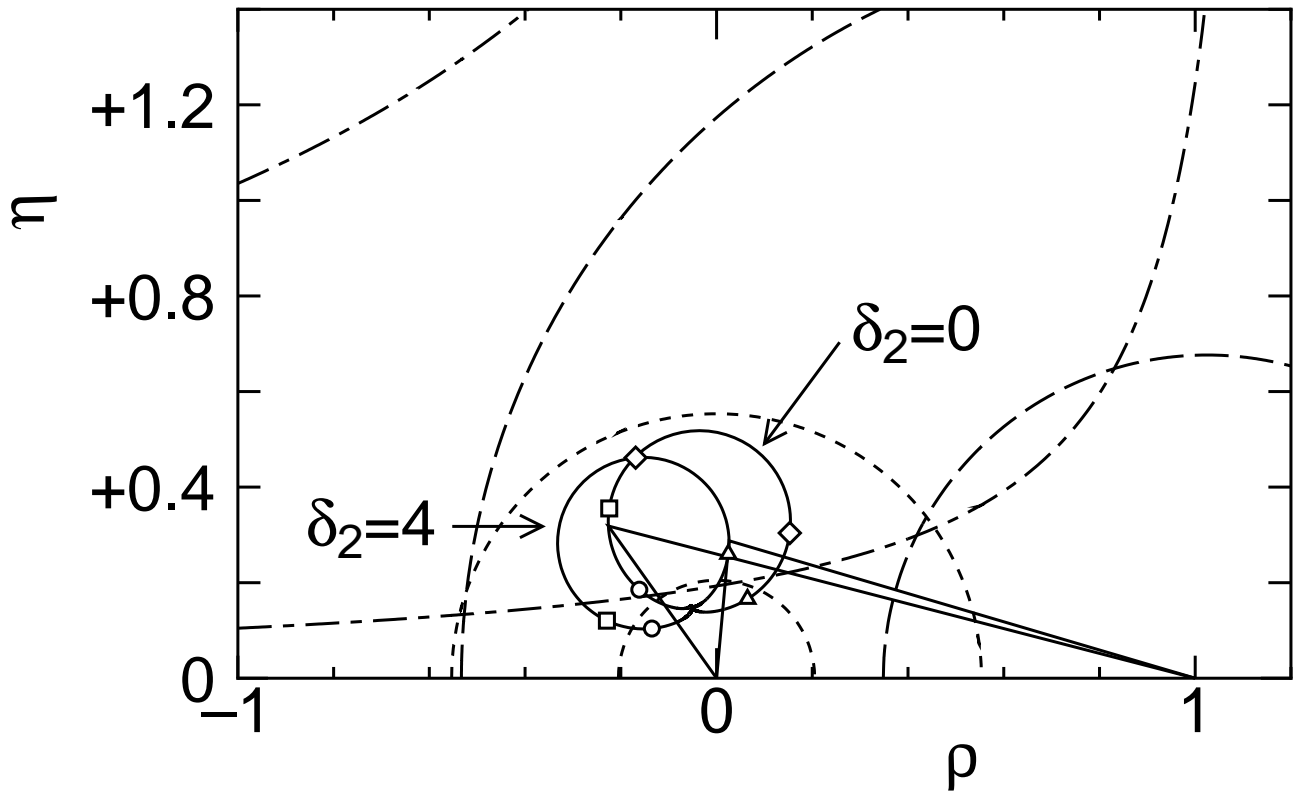
Fig.~3.  
Trajectories of the vertex $(\rho, \eta)$ of the unitary triangle 
for the cases  $\delta_2=0^\circ$ and $\delta_2=4^\circ$. 
The points $\circ$, $\Box$, $\Diamond$ and $\triangle$ denote the vertex 
$(\rho, \eta)$ for $\delta_3=180^\circ$, $190^\circ$, $200^\circ$ 
and $210^\circ$, respectivly.
The other parameters are fixed to $\kappa =10$, $\kappa/\lambda=0.02$, 
$b_u=-1/3$, $\beta_u=0$, 
$b_d=-1$, and $\beta_d=-18^\circ$ from the observed quark mass ratios.
The solid, broken and dot-dashed lines denote constraints from 
$|V_{ub}/V_{cb}|$, $|\Delta m_{B_d}|$ and $\varepsilon_K$.
The two triangles correspond to the cases 
$(\delta_2, \delta_3)=(0^\circ, 186^\circ)$ and $(4^\circ, 208^\circ)$, 
respectively.

\end{document}